\documentclass[%
reprint,
twocolumn,
superscriptaddress
]{revtex4-1}
\usepackage{amsmath}
\usepackage{mathtools} 
\usepackage{graphicx}
\usepackage{dcolumn}
\usepackage{bm}
\usepackage{hyperref}
\hypersetup{
	bookmarks=true,
	colorlinks=true,
	urlcolor=blue,
	citecolor=blue,
	linkcolor=blue, 
	pdftex,
	linktocpage=true, 
	linktoc=all,     
	hyperindex=true
}
\usepackage[mathlines]{lineno}

\usepackage{siunitx} 
\usepackage{comment} 
\usepackage{braket} 
\usepackage{xcolor}

\usepackage{newfloat}
\DeclareFloatingEnvironment[name={Supplementary Figure},fileext=lsf,listname={List of Supplementary Figures}]{suppfigure}

\usepackage[normalem]{ulem}
\newcommand\redout{\bgroup\markoverwith
	{\textcolor{red}{\rule[.5ex]{2pt}{0.4pt}}}\ULon}

\begin{document}


\title{Density Fluctuations across the Superfluid-Supersolid Phase Transition in a Dipolar Quantum Gas}%
\author{J. Hertkorn}%
\thanks{These authors contributed equally to this work.}
\affiliation{5. Physikalisches Institut and Center for Integrated Quantum Science and Technology, Universität Stuttgart, Pfaffenwaldring 57, 70569 Stuttgart, Germany
}%
\author{J.-N. Schmidt}%
\thanks{These authors contributed equally to this work.}
\affiliation{5. Physikalisches Institut and Center for Integrated Quantum Science and Technology, Universität Stuttgart, Pfaffenwaldring 57, 70569 Stuttgart, Germany
}%
\author{F. B\"ottcher}%
\affiliation{5. Physikalisches Institut and Center for Integrated Quantum Science and Technology, Universität Stuttgart, Pfaffenwaldring 57, 70569 Stuttgart, Germany
}%
\author{M. Guo}%
\affiliation{5. Physikalisches Institut and Center for Integrated Quantum Science and Technology, Universität Stuttgart, Pfaffenwaldring 57, 70569 Stuttgart, Germany
}%
\author{M. Schmidt}%
\affiliation{5. Physikalisches Institut and Center for Integrated Quantum Science and Technology, Universität Stuttgart, Pfaffenwaldring 57, 70569 Stuttgart, Germany
}%
\author{K.S.H. Ng}%
\affiliation{5. Physikalisches Institut and Center for Integrated Quantum Science and Technology, Universität Stuttgart, Pfaffenwaldring 57, 70569 Stuttgart, Germany
}%
\author{S.D. Graham}%
\affiliation{5. Physikalisches Institut and Center for Integrated Quantum Science and Technology, Universität Stuttgart, Pfaffenwaldring 57, 70569 Stuttgart, Germany
}%
\author{H.P. B\"uchler}%
\affiliation{Institute for Theoretical Physics III and Center for Integrated Quantum Science and Technology, Universität Stuttgart, Pfaffenwaldring 57, 70569 Stuttgart, Germany
}%
\author{T. Langen}%
\affiliation{5. Physikalisches Institut and Center for Integrated Quantum Science and Technology, Universität Stuttgart, Pfaffenwaldring 57, 70569 Stuttgart, Germany
}%
\author{M. Zwierlein}%
\affiliation{MIT-Harvard Center for Ultracold Atoms, Research Laboratory of Electronics, and Department of Physics, Massachusetts Institute of Technology, Cambridge, Massachusetts 02139, USA}
\author{T. Pfau}%
\email{t.pfau@physik.uni-stuttgart.de}
\affiliation{5. Physikalisches Institut and Center for Integrated Quantum Science and Technology, Universität Stuttgart, Pfaffenwaldring 57, 70569 Stuttgart, Germany
}%

\date{\today}

\begin{abstract}
	Phase transitions share the universal feature of enhanced fluctuations near the transition point. Here we show that density fluctuations reveal how a Bose-Einstein condensate of dipolar atoms spontaneously breaks its translation symmetry and enters the supersolid state of matter -- a phase that combines superfluidity with crystalline order. We report on the first direct in situ measurement of density fluctuations across the superfluid-supersolid phase transition. This allows us to introduce a general and straightforward way to extract the static structure factor, estimate the spectrum of elementary excitations and image the dominant fluctuation patterns. We observe a strong response in the static structure factor and infer a distinct roton minimum in the dispersion relation. Furthermore, we show that the characteristic fluctuations correspond to elementary excitations such as the roton modes, which have been theoretically predicted to be dominant at the quantum critical point, and that the supersolid state supports both superfluid as well as crystal phonons.
\end{abstract}

\maketitle

Fluctuations play a central role in quantum many-body systems. They connect the response and correlation of the system to its excitation spectrum, instabilities, phase transitions and thermodynamic properties.
A quantity that is fundamental to the theoretical description of fluctuations in many-body systems is the structure factor, which can be formulated as the Fourier transform of the density-density correlation function \cite{Griffin1993book,PitaevskiiBook2016}. Superfluid helium is an important example of a quantum many-body state, where the determination of the structure factor was crucial to understand its elementary excitations and therefore improved our understanding of superfluidity \cite{Feynman1954heTwoFluid,Tarvin1977,Balibar2007,PitaevskiiBook2016}. In the case of quantum gases, the structure factor of Bose-Einstein condensates (BECs) and superfluid Fermi gases is often investigated by Bragg spectroscopy \cite{Stenger1999,Veeravalli2008,Kuhnle2010}. In contact-interacting BECs this enabled the study of the spectrum and collective modes \cite{Stamper-Kurn1999}. In dipolar BECs, it has provided indications of the roton minimum in the dispersion relation \cite{Petter2019} analogous to the neutron and X-ray scattering data for helium \cite{Ceperley1995,PitaevskiiBook2016}. A different approach is to look at the condensate density directly in situ, which provides access to finite temperature and quantum fluctuations \cite{Folling2005,Esteve2006,Gemelke2009,Hung2011,Hung2011invariance,Hung2013,Schemmer2018}, and enables to extract the static structure factor simultaneously at all momenta.

The roton minimum both in helium and dipolar quantum gases is accompanied by a characteristic peak in the static structure factor close to the roton momentum \cite{Klawunn2011,PitaevskiiBook2016,Hofmann2020,Pal2020numberfluct}. Unlike in helium however, the contact interactions in dipolar quantum gases are tunable \cite{Chin2010}. This tunability allows for precise control of the dispersion relation and the controllable softening of the roton minimum. The roton modes associated to this minimum manifest as density modulations on top of the ground-state density distribution \cite{Santos2003, Giovanazzi2004}. An instability in the ground state can appear once the roton minimum is sufficiently soft. Since these modes represent precursors to solidification, dipolar BECs have long been proposed as candidates for the elusive supersolid state of matter, which simultaneously combines crystalline order with superfluidity \cite{Boninsegni2012}.

Recently, a dipolar supersolid state of matter has been realized through a phase transition from a BEC to an array of coherent quantum droplets \cite{Tanzi2019, Bottcher2019, Chomaz2019, Guo2019, Tanzi2019a, Natale2019, Bottcher2020} by precisely tuning the contact interaction strength. Close to the transition point these droplets are immersed in a superfluid background, and by lowering the scattering length further the superfluid fraction decreases towards a regime of isolated droplets. As the superfluid-supersolid phase transition is governed by intrinsic interactions it is of interest to study the fluctuations that emerge across the transition \cite{Zurek1996,Campo2013Causality,Pal2020numberfluct}, facilitate structure formation \cite{Hung2011,Mottl2012,Landig2015,Lozovik2020,Hofmann2020} and give rise to the supersolid state. The low-lying collective modes were shown to be particularly interesting regarding the aspect of supersolidity in this system \cite{Guo2019,Tanzi2019a,Natale2019,Hertkorn2019}. Those modes are facilitated by a continuous superfluidity across the droplet array, despite the translational symmetry breaking.

\onecolumngrid\
\begin{center}\
\begin{figure}[tb!]
	\includegraphics[trim=0 0 0 0,clip,scale=0.75]{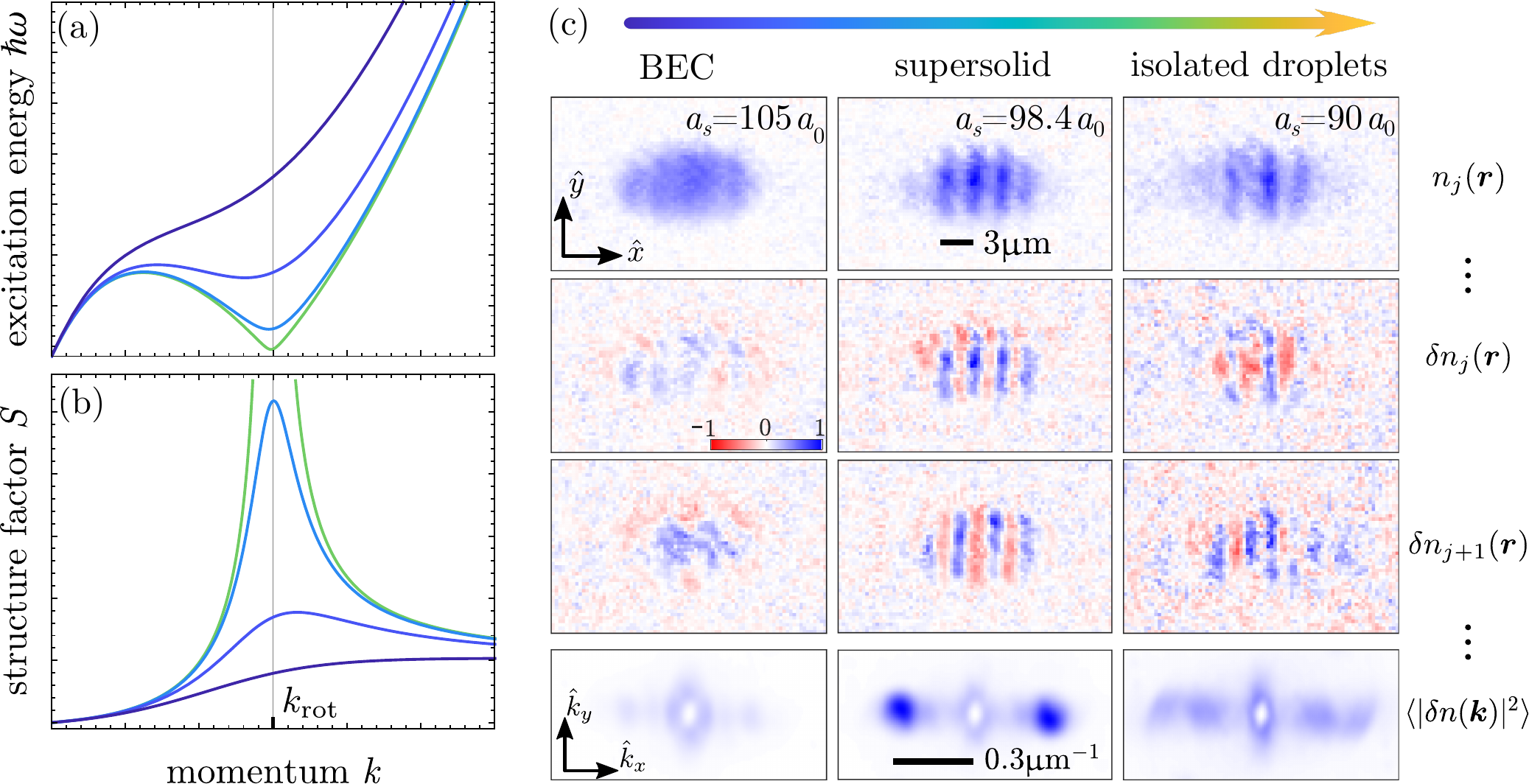}
	\caption{\textbf{Connection between rotonic dispersion relation, the static structure factor and the observed density fluctuations.} (a)-(b) Schematic of the dispersion relation $\hbar \omega(k)$ and associated static structure factor $S(k)$ of an elongated and strongly dipolar BEC \cite{Pal2020numberfluct}. A decrease in scattering length $a_s$ causes a roton minimum to emerge in the excitation spectrum, associated with a characteristic peak in the static structure factor. The roton momentum $k_\mathrm{rot}$ is indicated where the roton minimum drops near zero. (c) For a given scattering length $a_s$ we observe a large number of in situ densities $n_j(\boldsymbol{r})$, calculate their mean $\braket{n(\boldsymbol{r})}$ and the density fluctuations $\delta n_j(\boldsymbol{r}) = n_j(\boldsymbol{r})-\braket{n(\boldsymbol{r})}$ as the deviation of the in situ images from their mean. Investigating the mean power spectrum of the fluctuations $\braket{\left| \delta n(\boldsymbol{k}) \right|^2}$ for different scattering lengths across the transition allows us to directly observe the static structure factor as the system passes from BEC to supersolid to isolated droplet states. The colored arrow on top indicates the direction towards lower scattering lengths, passing from BEC to supersolid to isolated droplet regimes. The colormap used for the images shows the normalized amplitude of densities, density fluctuations and mean power spectra, respectively.}
	\label{fig:schematic}
\end{figure}\
\end{center}\
\twocolumngrid\
\noindent
Here we provide the first direct in situ observation of density fluctuations across the superfluid-supersolid phase transition in a trapped dipolar quantum gas. By analyzing hundreds of in situ images of the atomic cloud around the phase transition point we spatially resolve characteristic fluctuation patterns that arise across the transition. From the observed fluctuations we determine the static structure factor and estimate the spectrum of elementary excitations. We observe a strong peak in the static structure and an associated roton minimum in the dispersion relation. Moreover, we experimentally determine that the dominant fluctuations at the transition point correspond to two degenerate roton modes \cite{Guo2019, Hertkorn2019} and that the supersolid state supports both superfluid as well as crystal phonons in a narrow range of scattering lengths. Our study combines the fluctuations with the excitation spectrum of a dipolar supersolid and highlights its bipartite nature between the superfluid BEC and the crystalline isolated droplets. 

The rotonic dispersion relation of strongly dipolar BECs \cite{Santos2003, Giovanazzi2004, Petter2019} is schematically shown in Fig.~\ref{fig:schematic}(a). The system becomes more susceptible to density fluctuations as the roton minimum softens. These density fluctuations are associated with a characteristic peak \cite{Arkhipov2005,Astrakharchik2007,Blakie2012,Blakie2013depletionFluctuations,PitaevskiiBook2016} in the static structure factor, illustrated in Fig.~\ref{fig:schematic}(b). This can be understood by considering the general Feynman-Bijl formula ${S(\boldsymbol{k})=\hbar^2 \boldsymbol{k}^2/2m\varepsilon(\boldsymbol{k})}$ \cite{Feynman1954heTwoFluid,PitaevskiiBook2016}, connecting the static structure factor $S(\boldsymbol{k})$ to the excitation spectrum $\epsilon(\boldsymbol{k})$ at zero temperature. As the energy of the roton modes drops near zero, the density fluctuations and thus the structure factor increase dramatically. Eventually, the roton minimum has sufficiently softened in order for the system to enter the roton instability. This roton instability triggers the phase transition to a dipolar supersolid and arrays of isolated quantum droplets. 

To study the static structure factor experimentally, we prepare a dipolar BEC with typically ${40 \times 10^3}$ $^{162}$Dy atoms at temperatures of approximately $20\,\si{\nano\kelvin}$ in a cigar-shaped trap with trapping frequencies ${\omega/2\pi= [30(1),\,89(2),\,108(2)]\,\si{\hertz}}$ and a magnetic field oriented along $\hat{\boldsymbol{y}}$ (see Methods). The scattering length is tuned via a double Feshbach resonance \cite{Bottcher2019droplet} to final values between $90\,a_0$ and $105\,a_0$ by linearly ramping the magnetic field in $30\,\si{\milli\second}$. We wait for $15\,\si{\milli\second}$ to allow for the system to equilibrate and then the atoms are imaged using phase-contrast imaging along the $\hat{\boldsymbol{z}}$-axis with a resolution of $\sim 1\,\si{\micro\meter}$. We find either a BEC, a supersolid phase (SSP) or isolated droplets (ID) for large ($a_s \simeq 105\,a_0$), intermediate  ($a_s \simeq 98.4\,a_0$) and small ($a_s \simeq 90\,a_0$) scattering lenghts, respectively \cite{Guo2019}. We accumulate enough averages for a statistical evaluation of the structure factor by repeating the experiment around $200$ times for every scattering length.

We obtain $S(\boldsymbol{k})$ experimentally by analyzing the in situ images as illustrated in Fig.~\ref{fig:schematic}(c). For every scattering length we center the in situ densities  $n_j(\boldsymbol{r})$ to their center of mass and normalize them to the mean atom number. With the former step we remove contributions of the dipole center of mass motion \cite{Guo2019} and with the latter we correct for shot to shot total atom number fluctuations (\cite{Esteve2006}; Methods) that would otherwise give contributions to $S(\boldsymbol{k})$ near $\boldsymbol{k} = 0$. From these in situ images we obtain the mean image $\braket{n(\boldsymbol{r})}$ and the density fluctuations $\delta n_j(\boldsymbol{r}) = n_j(\boldsymbol{r})-\braket{n(\boldsymbol{r})}$ as the deviation of the in situ images from their mean. With the Fourier transform of the density fluctuation $\delta n_j (\boldsymbol{k}) = \int\!\mathrm{d}^3r\delta n_j (\boldsymbol{r})e^{i \boldsymbol{k}\cdot \boldsymbol{r}}$ we obtain the mean power spectrum of the fluctuations $\braket{\left| \delta n(\boldsymbol{k}) \right|^2}$. In homogeneous systems, the static structure factor can be directly written as $S(\boldsymbol{k}) = \braket{\left|\delta n(\boldsymbol{k})\right|^2}/N$, where $N$ is the atom number \cite{Hung2011,Hung2013,PitaevskiiBook2016}.
In practice, the interpretation is less straightforward \cite{Naraschewski1999,Zambelli2000,Steinhauer2004,Schemmer2018} since the expectation values of the density $\braket{n(\boldsymbol{r})}$ are spatially dependent due to the finite size and the translational symmetry breaking in the supersolid and droplet regime. Nonetheless $S(\boldsymbol{k})$ gives insight into the strength of fluctuations \cite{Esteve2006,Imambekov2009,Armijo2010,Jacqmin2011,Schemmer2018} and is a quantity that can be continuously evaluated from the BEC via the supersolid to the isolated droplet regime. We note that our evaluation is limited to intermediate momenta between $k_\mathrm{min} / 2 \pi \simeq 0.08\,\si{\micro\meter\tothe{-1}}$  and $k_\mathrm{max} / 2 \pi \simeq 1\,\si{\micro\meter\tothe{-1}}$ due to the finite system size and the finite resolution of our imaging system, respectively \cite{Hung2011,Schemmer2018}. We extract the static structure factor $S(k_x,k_y,k_z=0)$ cut along the $k_z=0$ plane according to the Fourier slice theorem since the atomic densities are integrated along the line of sight during the imaging process (\cite{Blumkin2013}; Methods). Due to the cigar-shaped trap geometry, the fluctuations predominantly show structure along $\hat{k}_x$ (see Fig.~\ref{fig:schematic}(c)), which allows us to extract a cut of the mean power spectrum at $k_y=0$ to obtain the 1D structure factor $S(k_x)$.

\begin{figure}[tb!]
	\includegraphics[trim=0 0 0 0,clip,scale=0.44]{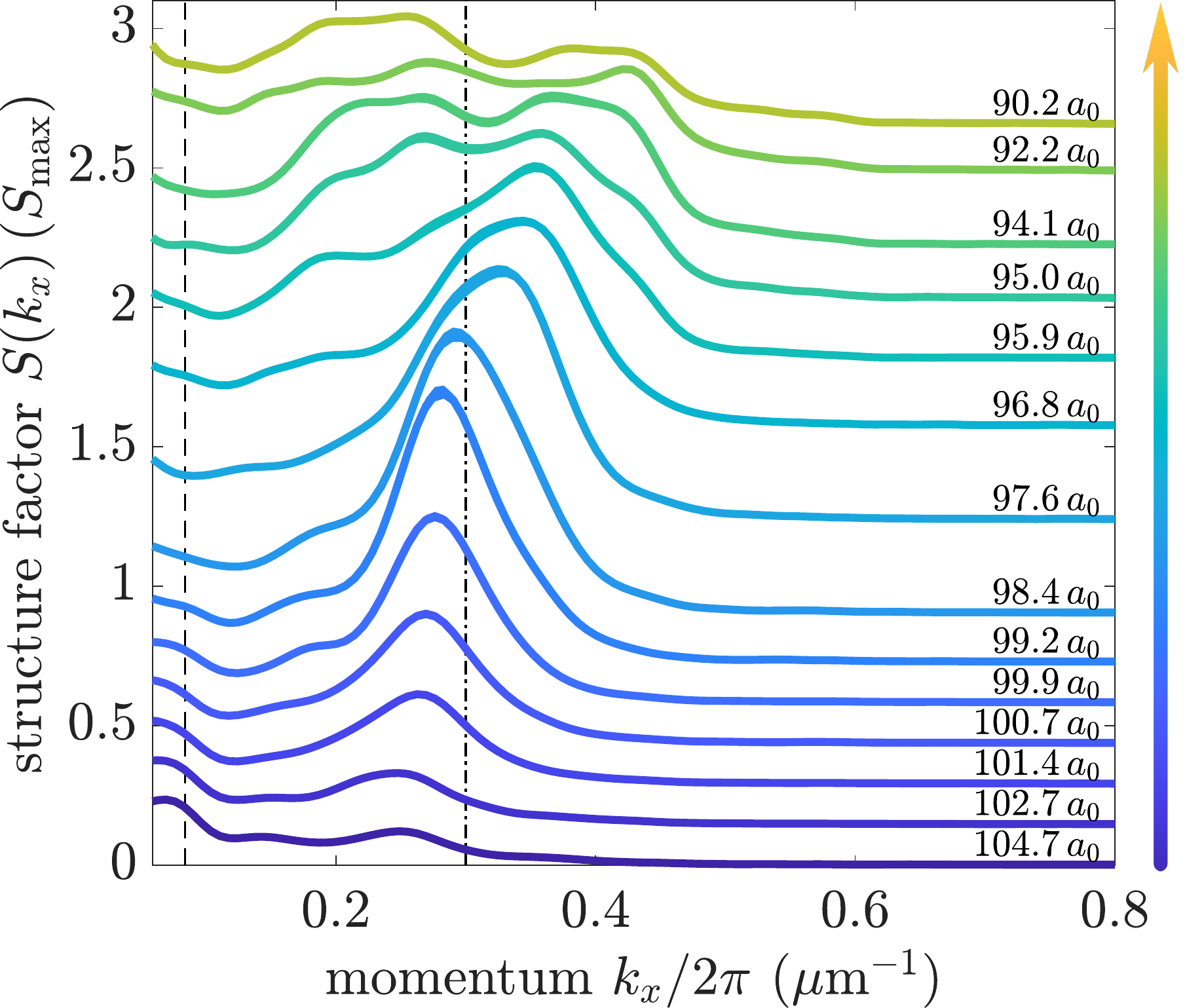}
	\caption{\textbf{Static structure factor across the transition.}  Experimentally determined structure factor in units of $S_{\mathrm{max}}= 260$ for different scattering lengths $a_s$. The dashed line on the left indicates the smallest momentum ${k_\mathrm{min}/2\pi \simeq 0.08\,\si{\micro\meter\tothe{-1}}}$ available due to the finite size of the system along the $x$-direction. The dash-dotted line at ${k_x/2\pi \simeq 0.3\,\si{\micro\meter\tothe{-1}}}$ roughly indicates the inverse droplet spacing in the droplet regime, which coincides with the roton momentum at the transition point. Errors are indicated as increased thickness of the lines and are obtained by the bootstrapping method (\cite{Efron1979}; Methods). For illustration purposes the lines for smaller $a_s$ have been shifted up. The momentum axis is sampled with a spacing of ${\Delta k_x / 2\pi \simeq 0.007\,\si{\micro\meter^{-1}}}$. The colored arrow indicates the direction towards lower scattering lengths, passing from BEC to supersolid to isolated droplet regimes.}
	\label{fig:waterfall}
\end{figure}

Using the above described analysis, we obtain $S(k_x)$ across the phase transition as shown in Fig.~\ref{fig:waterfall}. In the BEC regime at $a_s \simeq 104\,a_0$, we find the structure factor to be relatively flat with the exception of a small peak at around $k_x/2\pi \simeq 0.25\,\si{\micro\meter\tothe{-1}}$. This peak is an indication that far in the BEC regime roton modes can be excited \cite{Guo2019} and consequently that the spectrum features modifications from a purely contact interacting quantum gas. 

As the scattering length is reduced, the position and amplitude of this characteristic peak are observed to increase continuously towards the phase transition point ($a_s \simeq 98.4\,a_0$). A continuously growing peak amplitude of the structure factor signals enhanced fluctuations, consistent with a softening roton minimum towards the transition point. The structure factor reaches its maximum value as a function of the scattering length at the transition point and is located at the roton momentum $k_\mathrm{rot}/2\pi \simeq 0.29\,\si{\micro\meter\tothe{-1}}$. This value is mainly given by the harmonic oscillator length $l_y$ along the magnetic field direction \cite{Chomaz2018}. At the transition point, the enhanced fluctuations provide the roton instability and lead to the formation of supersolid quantum droplets, whose spacing $d \simeq 3\,\si{\micro\meter}$ smoothly matches the roton wavelength $2\pi/k_\mathrm{rot}$ \cite{Bottcher2019,Guo2019}. The observed increase of the roton momentum towards smaller scattering lengths can be understood in a variational approach of elongated dipolar condenstates \cite{Blakie2020variational}. Around the transition point, the in situ densities we observe from shot to shot show droplets immersed in an overall BEC background, constituting the supersolid state of matter \cite{Tanzi2019, Bottcher2019, Chomaz2019, Guo2019, Tanzi2019a, Natale2019, Roccuzzo2019, Hertkorn2019, Bottcher2020}. Here the density fluctuation patterns become most clear and show spatial oscillations (see Fig.~\ref{fig:schematic}(c), middle column). These characteristic fluctuations can directly be attributed to the symmetric and antisymmetric roton modes we found in our previous work \cite{Hertkorn2019}.

After crossing the phase transition ($a_s \lesssim 98.4\,a_0$) the peak amplitude of the structure factor decreases and a shoulder develops at smaller momenta. This shoulder increases further for smaller scattering lengths and eventually leads to a double-peak structure as seen in Fig.~\ref{fig:waterfall}. The origin of this rising double peak can be understood by means of a principle component analysis.

The maximum of the structure factor for different scattering lengths acts as a measure of the density fluctuation strength across the superfluid to supersolid phase transition. It quickly increases from the BEC side when approaching the phase transition indicating a significant enhancement of the characteristic fluctuations close to the phase transition point.  We see that the increase from the BEC side towards the phase transition is sharper than the decrease on the droplet side. The magnitude of the structure factor ($S_{\mathrm{max}} = 260$) can mainly be explained by thermal enhancement of the participating low-energy modes.

\begin{figure}[tb!]
	\includegraphics[trim=0 0 0 0,clip,scale=0.5]{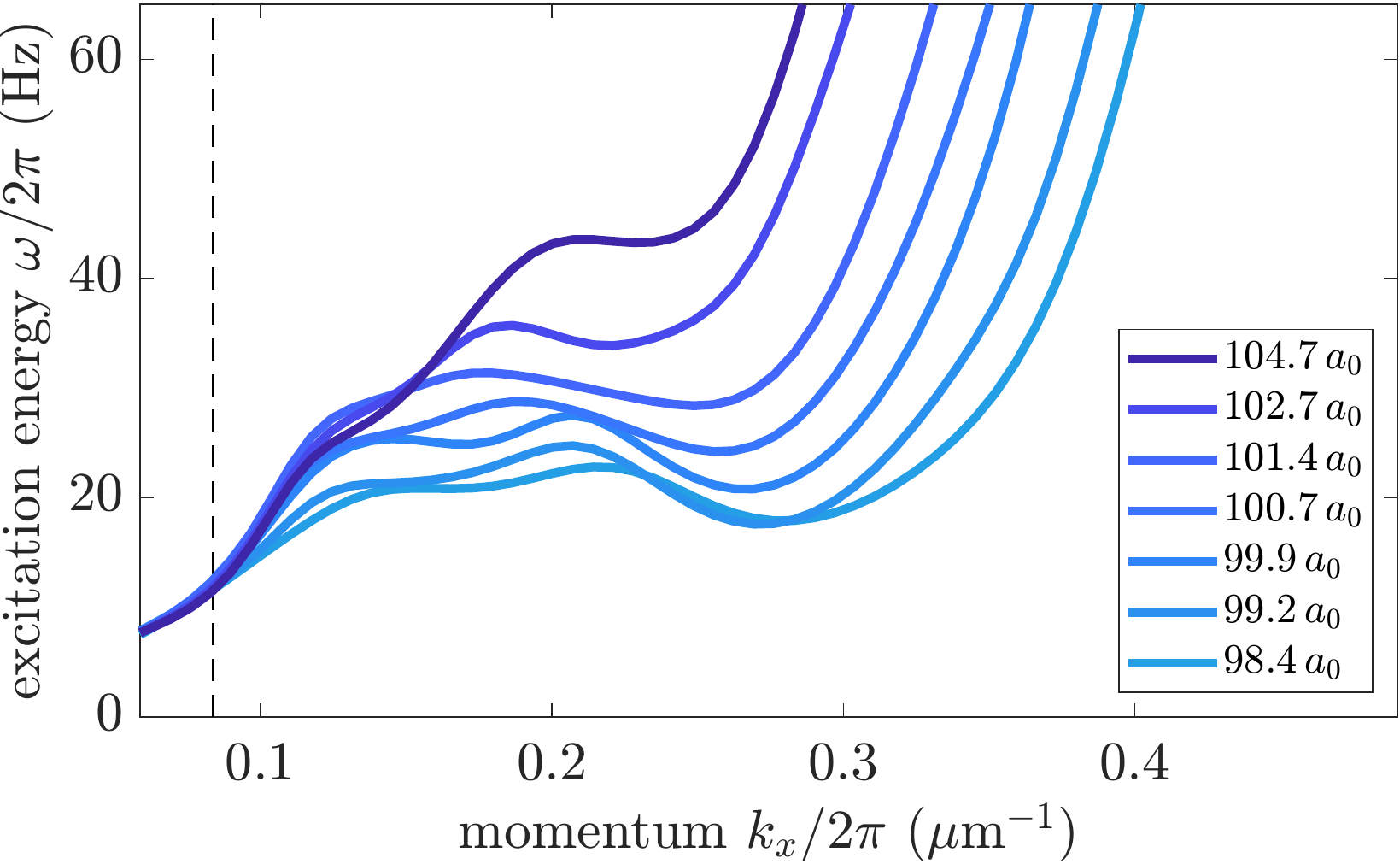}
	\caption{\textbf{Dispersion relation from BEC to supersolid.} Experimentally determined excitation energy $\omega(k_x)$ according to equation~\eqref{eq:StrucTemp} assuming a temperature of \SI{20}{\nano\kelvin}, for scattering lengths above  the phase transition point. A clear roton minimum at finite momentum  is observed that softens towards the transition point.}
	\label{fig:Dispersion}
\end{figure}

To estimate the dispersion relation based on the experimentally determined structure factor we use the relation
\begin{equation}\label{eq:StrucTemp}
S(\boldsymbol{k}) = \frac{\hbar^2 \boldsymbol{k}^2}{2m\varepsilon(\boldsymbol{k})} \coth\left(\frac{\varepsilon(\boldsymbol{k})}{2k_\mathrm{B}T}\right),
\end{equation}
which extends the Feynman-Bijl formula, ${S(\boldsymbol{k})=\hbar^2 \boldsymbol{k}^2/2m\varepsilon(\boldsymbol{k})}$, valid at $T = 0$, to nonzero temperatures $T$ \cite{Zambelli2000,PitaevskiiBook2016}. At nonzero temperatures and small excitation energies the contribution of low-lying modes to the structure factor can easily be enhanced by several orders of magnitude. Close to the transition point where the roton gap $\Delta_\mathrm{rot}$ is small compared to the temperature of the system ($\hbar\Delta_\mathrm{rot}/k_\mathrm{B}T \lesssim 1$), equation~\eqref{eq:StrucTemp} can be expanded and the peak of the static structure factor scales as $S_\mathrm{max} \sim T/\Delta_\mathrm{rot}^2$ \cite{Klawunn2011}. Note that equation~\eqref{eq:StrucTemp} is an excellent description of the structure factor for a weakly interacting superfluid, where the excitation spectrum is dominated by a single mode, and where the influence of the quantum as well as thermal depletion can be ignored. Although we study a finite system, leading to a discrete excitation spectrum, a continuous approximation to the dispersion relation yields a meaningful estimate for the excitation energies (see Methods). We show the resulting spectrum ${\omega(k_x)/2\pi = \varepsilon(k_x)/h}$ in Fig.~\ref{fig:Dispersion}. To do so we assumed a mean temperature of \SI{20}{\nano\kelvin}, a conservative approximation to include additional minor heating during the preparation (see Methods). In Fig.~\ref{fig:Dispersion}, one can see a small roton minimum already well above the trap frequency of \SI{30}{\hertz} for a large scattering length. The roton minimum softens and moves towards higher momenta $k_x$ as the scattering length is lowered and finally reaches its lowest energy at the phase transition point. After crossing the phase transition point, when the crystalline structure has developed, the excitation spectrum should have a band structure due to the translational symmetry breaking. In this case equation~\eqref{eq:StrucTemp} is no longer necessarily justified as several modes contribute to the excitation spectrum, and therefore it is no longer straightforward to extract the excitation spectrum from the measured static structure factor.

\begin{figure}[tb!]
	\includegraphics[trim=0 0 0 0,clip,scale=0.525]{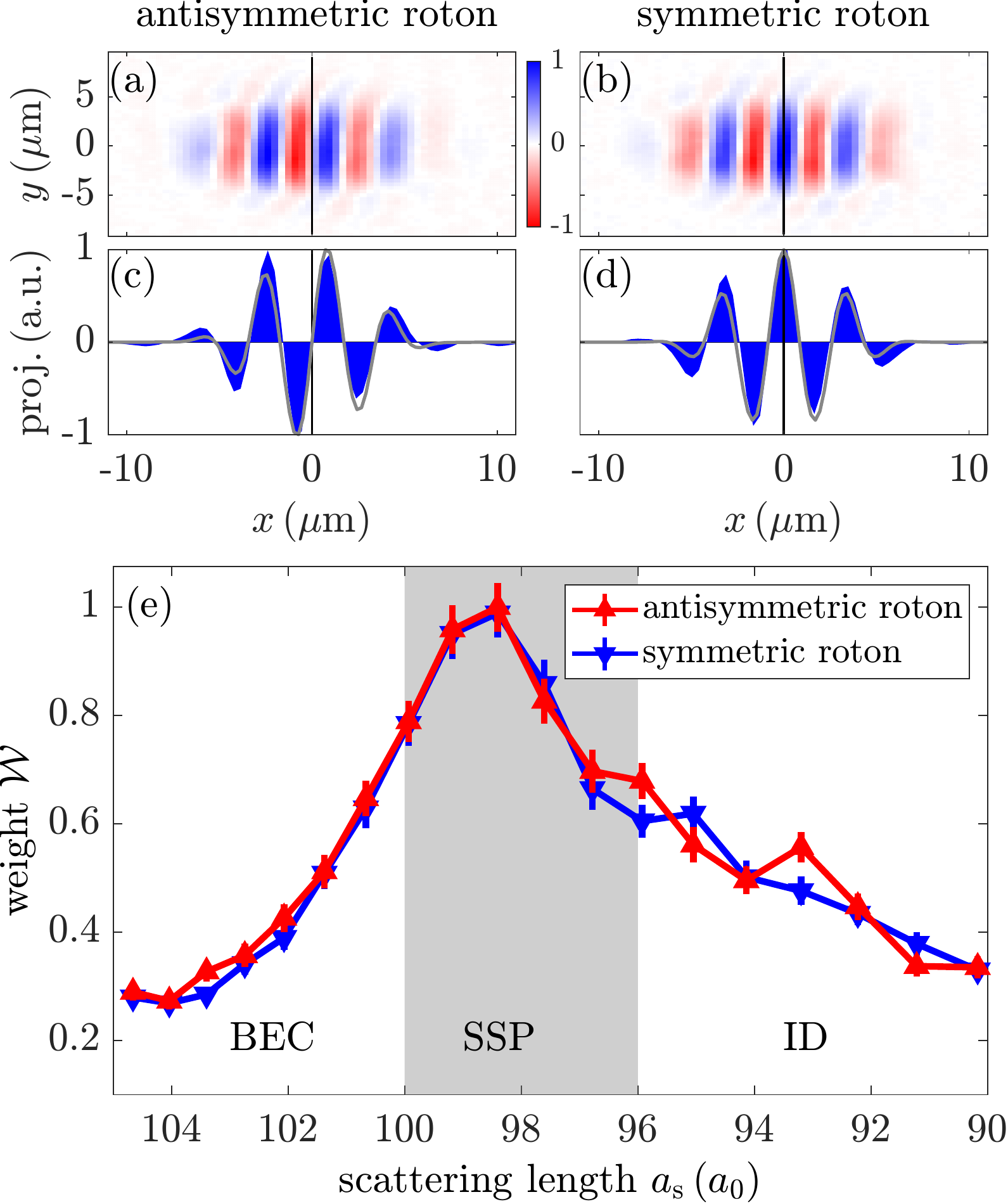}
	\caption{\textbf{Roton modes and their weight across the transition.} (a)-(b) Spatial structure of two principal components with the largest weight dominating the experimental dataset. (c)-(d) Projections of (a)-(b) onto the $x$-axis (blue), with a comparison to the antisymmetric and symmetric roton density fluctuations from our BdG calculation \cite{Hertkorn2019} (gray). (e) Mean absolute weight $\mathcal{W}$ of the symmetric and antisymmetric roton, normalized to the weight of the antisymmetric roton mode at the transition point, which is the maximum weight of all principle components over the whole scattering length range. The roton modes are dominant and their weights are almost degenerate from the BEC ($a_s = 104.7\,a_0$) leading up to the transition point ($a_s = 98.4\,a_0$). The gray area indicates the supersolid region previously determined \cite{Guo2019}. Error bars indicate the standard error of the mean. }
	\label{fig:pca}
\end{figure}

\begin{figure}[tb!]
	\includegraphics[trim=0 0 0 0,clip,scale=0.575]{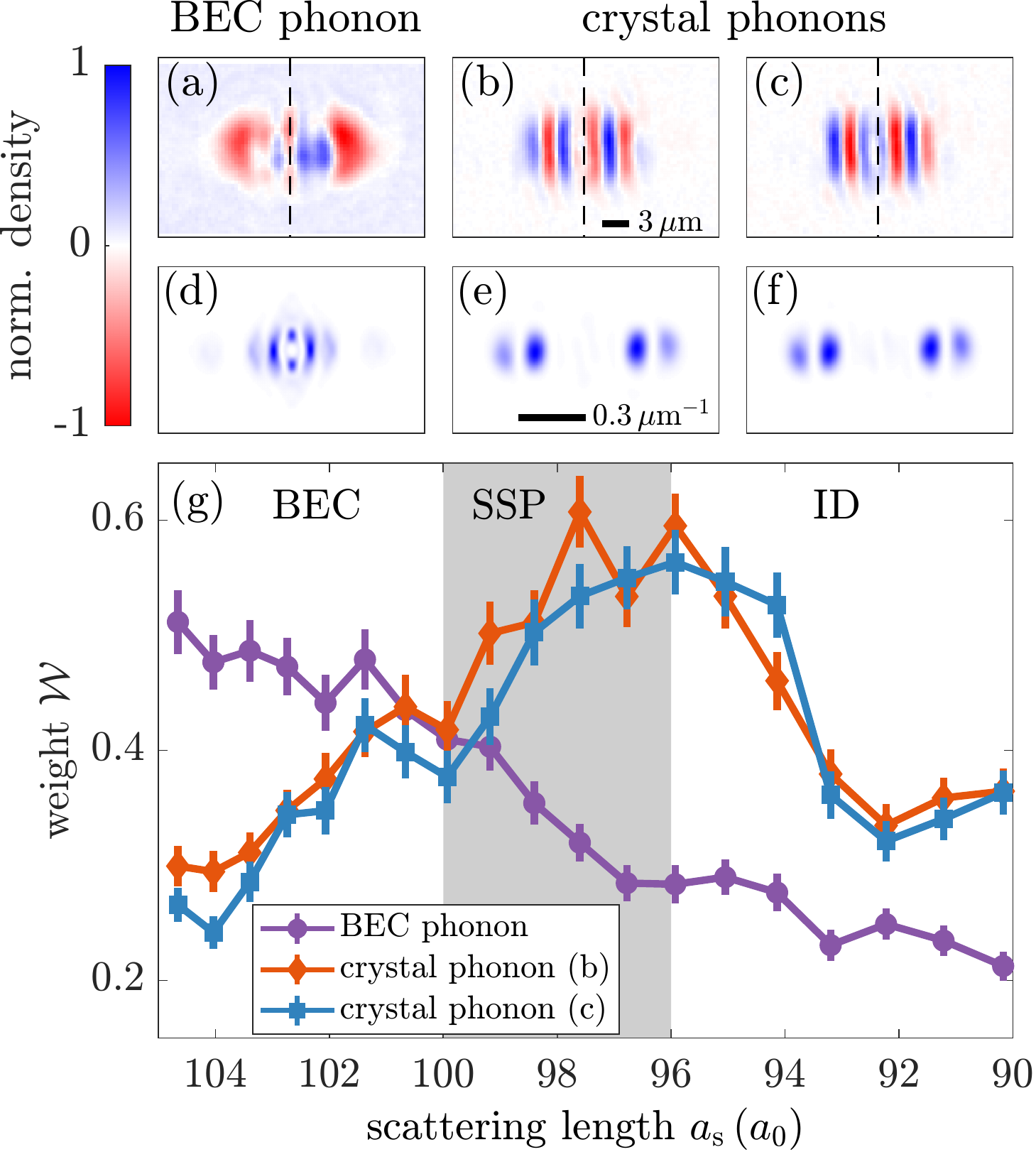}
	\caption{\textbf{Signature of BEC and crystal phonons.} Higher principle components corresponding to the phonons of the crystal and the BEC (a)-(c) including their Fourier transforms (d)-(f) and weights (g). (a) Quadrupole mode of the BEC. The antisymmetric (b) and the symmetric (c) breathing mode of the crystal show a clear splitting in their Fourier transform (e)-(f). (g) Comparison of the mean absolute weights $\mathcal{W}$ of these modes featuring a clear overlap region that indicates a supersolid phase. The gray shaded area shows the previously determined supersolid region \cite{Guo2019}. Error bars indicate the standard error of the mean. The weights are normalized to the weight of the antisymmetric roton mode at the transition point, which is the maximal weight of all principle components over the whole scattering length range.}
	\label{fig:pcacolor}
\end{figure}

To gain a better insight into the modes that dominantly contribute to the fluctuations, we use principal component analysis (PCA) \cite{Jolliffe2002} on the density fluctuations for all scattering lengths combined. This model-free statistical analysis is a general method to extract dominant components or to reduce the dimensionality of a dataset. We study the principal components (PCs) across the phase transition since there is a direct correspondence \cite{Dubessy2014} to the dominant collective excitations obtained with the Bogoliubov-de-Gennes (BdG) formalism \cite{Hertkorn2019}. This allows us to identify and compare the most dominant PCs with specific BdG modes and study how their weight behaves across the transition, as shown in Fig.~\ref{fig:pca} and Fig.~\ref{fig:pcacolor}.

The first principal component is structureless and only represents the global atom number fluctuation \cite{Dubessy2014,Barr2015}. The subsequent principal components are shown in Fig.~\ref{fig:pca}(a)-(b). These two components are dominant across the phase transition and represent a periodic spatial pattern. Close to the transition point we can identify these characteristic patterns in many single-shot realizations of the density fluctuations, as shown in the central column of Fig.~\ref{fig:schematic}(c). We compare the profiles of these PCs to the antisymmetric and symmetric roton modes from the BdG theory at the transition point in Fig.~\ref{fig:pca}(c)-(d) and find them to be in excellent agreement. These two roton modes are developing into the Goldstone \cite{Guo2019} and amplitude (Higgs) \cite{Hertkorn2019} modes of the supersolid.

The mean absolute value of the weights for these two PCs are shown in Fig.~\ref{fig:pca}(e) as an indication of their strength across the phase transition. Starting from the BEC side, where they are comparable in strength to other modes, these PCs gain rapidly in strength as the phase transition is approached (see Methods). We note that the maximum of the structure factor behaves similar to the weights of these two PCs as a function of scattering length. Leading up to the quantum critical point at $a_s \simeq 98.4\, a_0$ these two modes have identical weights, in accordance to our previous work \cite{Hertkorn2019}, in which we showed that the two roton modes remain degenerate while softening towards the phase transition.

Further into the isolated droplet regime the weight of the roton PCs decrease and other PCs become more important because further modes are softening. In Fig.~\ref{fig:pcacolor} we present the three next higher PCs that correspond to the BEC phonon (a) and the antisymmetric (b) and symmetric (c) crystal phonon, respectively. The quadrupole mode of the BEC has a relatively high weight in the BEC regime and vanishes for small scattering lengths towards the isolated droplet regime. The breathing or lowest phonon modes of the droplet crystal show a clear splitting in the Fourier transform (Fig.~\ref{fig:pcacolor}(d)-(e)) explaining the observed double-peak structure in $S(k_x)$ for low scattering lengths. This can be understood as the appearance of the band structure, where excitations are split around the edge of the Brillouin zone. These modes only have an appreciable weight for low scattering length after crossing the phase transition (Fig.~\ref{fig:pcacolor}(g)). In the experiment the excitation of the crystal breathing mode is further enhanced by the preparation process \cite{Bottcher2019, Tanzi2019a}. Note that there is a small region close to the phase transition where both types of modes have a non-vanishing weight. This subtle feature shows the coexistence of both BEC and droplet crystal, which is a prerequisite of the supersolid nature of the phase. One can see that the supersolid state supports both types of excitations -- the phonon of the superfluid BEC and the crystalline phonons of the droplets. 

In conclusion, we reported the first in situ measurement of the density fluctuations across the superfluid to supersolid phase transition in a dipolar quantum gas. We quantified the fluctuation strength across the transition by the static structure factor $S(k_x)$ using a statistical evaluation of in situ images and found a characteristic peak in $S(k_x)$ that strongly increases towards the phase transition point. We showed that this peak is unambiguously dominated by the low-lying modes of the rotonic dispersion relation. The characteristic fluctuations close to the transition point are stronger compared to the BEC or isolated droplet regime. The large amplitude of the measured static structure factor reveals the important role played by temperature at the phase transition, an aspect which has so far been absent in the discussion of the superfluid-supersolid phase transition. Using principal component analysis we spatially resolved the dominant fluctuations and identified them as two roton modes. Furthermore, we showed that the supersolid supports both superfluid and crystal phonons.  Our study provides a promising outlook to extract thermodynamic properties \cite{Hofmann2020} and possibly universal access to the condensate fraction \cite{Lozovik2020} of the supersolid state. Exciting avenues for future work includes using fluctuations as a tool for thermometry of the supersolid state \cite{Zhou2011}, understanding the out of equilibrium dynamics that arise when crossing the phase transition \cite{Langen2015,Eisert2015}, and exploring the roles of fluctuations in the Kibble-Zurek mechanism \cite{Zurek1996,Campo2013,Campo2013Causality}.

\begin{acknowledgements}
We thank Tobias Ilg and Jan Kumlin for valuable discussions. M.G. and M.Z. acknowledge funding from the Alexander von Humboldt Foundation. This work is supported by the German Research Foundation (DFG) within FOR2247 under Pf381/16-1 and Bu2247/1, Pf381/20-1, FUGG INST41/1056-1 and the QUANT:ERA collaborative project MAQS.
\end{acknowledgements}

\begin{appendix}

\section*{Author contributions}
J.N.S., F.B., M.G. performed the experiment. J.H., J.N.S., F.B., M.G. analyzed the data. J.H. performed the numerical analysis. H.P.B., T.L., M.Z. and T.P. provided scientific guidance in experimental and theoretical questions. All authors contributed to the interpretation of the data and the writing of the manuscript.

\section*{Competing interests}
The authors declare no competing financial interests.

\section*{Data availability}
Correspondence and requests for materials should be addressed to T.P.~(email: t.pfau@physik.uni-stuttgart.de).

\end{appendix}

\bibliographystyle{apsrev4-1}
\bibliography{refs_cleaned} 

\clearpage

\section*{Methods}

\subsection{Experimental protocol}

The complete experimental procedure has been described in detail in our previous publications \cite{Kadau2016,Bottcher2019,Guo2019}. After preparing a quasipure BEC of $^{162}$Dy with ${T\simeq 10\,\si{\nano\kelvin}}$ in a crossed optical dipole trap at \SI{1064}{\nano\meter}, we reshape the trap within \SI{20}{ms} to its final geometry with trap frequencies of $\omega / 2 \pi = \lbrack 30(1),\,89(2),\,108(2)\rbrack\,\si{\hertz}$. The magnetic field is oriented along the $\hat{\boldsymbol{y}}$-axis and is used to tune the contact interaction strength. 

We ramp the magnetic field in a two-step ramp closer to the double Feshbach resonance near \SI{5.1}{G} to tune the scattering length from its initial background value of ${a_{\mathrm{bg}} = 140(20)\,a_0}$ \cite{Tang2015a, Tang2016, Tang2018} to a final value between ${90\,a_0}$ and ${105\,a_0}$. This corresponds to the droplet and BEC regime, respectively. We expect the preparation scheme to induce some additional heating and thus assume a temperature of \SI{20}{\nano\kelvin} in our later analysis of $S(k_x)$ and $\omega(k_x)$. After \SI{15}{\milli\second} of free evolution the droplets have formed and equilibrated. We finally probe the atomic system using in situ phase-contrast imaging along the vertical $\hat{\boldsymbol{z}}$-axis, which is done using a microscope objective featuring a numerical aperture of $0.3$. We reach an imaging resolution of \SI{1}{\micro\meter}.

\subsection{Analysis method and principle component analysis}
In the following we will describe our analysis procedure. We start with a large set of images that contains 200 averages for each scattering length. We center the images with respect to their center-of-mass to get rid of the otherwise dominating dipole modes. Afterwards, we post-select on atom number for every scattering length and only take images with an atom number that lays in an interval of $\pm 30 \%$ around the mean atom number at that scattering length. We have confirmed that changing the tolerance in the post selection does not affect the features of the structure factor. In a next step, we normalize each image to its atom number $\tilde{n}_j = n_j/N_j$ and calculate the fluctuations $\delta \tilde{n}_j(\boldsymbol{r}) = \tilde{n}_j(\boldsymbol{r}) - \braket{\tilde{n}(\boldsymbol{r})}$ as the deviation of the normalized image $\tilde{n}_j(\boldsymbol{r})$ from the mean image $\braket{\tilde{n}(\boldsymbol{r})}$. The structure factor is then given by the power spectrum of these fluctuations multiplied by the mean atom number $\bar{N}$

\begin{equation}\label{eq:Sfac}
S(\boldsymbol{k}) = \bar{N}\langle | \delta \tilde{n}(\boldsymbol{k}) |^2 \rangle,
\end{equation}

where $\delta \tilde{n}_j (\boldsymbol{k}) = \mathcal{F}[\delta \tilde{n}_j] (\boldsymbol{k}) = \int\!\mathrm{d}^2r\,\delta \tilde{n}_j (\boldsymbol{r})e^{i \boldsymbol{k}\cdot \boldsymbol{r}}$ is the Fourier transform of the normalized fluctuations which were obtained from the line integrated images. According to the Fourier-slice theorem, taking the Fourier transform of two-dimensional integrated atomic densities is connected with the Fourier transform of the full three-dimensional atomic density distribution. For an arbitrary function $f(x,y,z)$ and its projection $p(x,y)=\int\!\mathrm{d}z\, f(x,y,z)$ the Fourier-slice theorem reads

\begin{equation}\label{eqn:FourierSlice}
\mathcal{F}\lbrack f \rbrack (k_x,k_y,0) = \mathcal{F}\lbrack p \rbrack (k_x,k_y).
\end{equation}

As a result, the static structure factor $S(k_x,k_y)$ extracted from the integrated densities is in fact a slice through the structure factor $\bar{S}(k_x,k_y,k_z)$ one would get if one had access to the full three-dimensional density distribution $S(k_x,k_y)=\bar{S}(k_x,k_y,0)$. Further it is worth noticing that normalizing each image to its atom number when comparing the structure factor at different scattering lengths has an effect in the small $k$ regime. This part of the structure factor is highly dependent on the total atom number which is changing across the transition due to higher three-body losses towards the droplet regime (Fig.~(\ref{fig:atomnumber})). Finally we take a cut at $k_y/2\pi =0$ to determine the one-dimensional static structure factor $S(k_x)$ we present in Fig.~\ref{fig:waterfall} of the main text. We confirmed that averaging within the $k_y$-values that correspond to our imaging resolution is below our statistical error obtained by bootstrapping \cite{Efron1979,Efron1993book,Politis1994,Politis1999book}.

Two natural boundaries arise at small-$k$ and at large-$k$ because we are probing a finite system. The small-$k$ cut-off simply comes from the finite size of the atomic cloud on the image. This results in a lowest $k$-value ${k_{\mathrm{min}}/2\pi \simeq  0.08\,\si{\micro\meter\tothe{-1}}}$ that a possible excitation must have in a system of size $L\simeq 12\,\si{\micro\meter}$. In contrast, the high-$k$ cut-off has its origin in the finite imaging resolution of the microscope objective which leads to ${k_{\mathrm{max}}/2\pi \simeq  1\,\si{\micro\meter\tothe{-1}}}$.

\begin{suppfigure}[tb!]
	\includegraphics[trim=0 0 0 0,clip,scale=0.37]{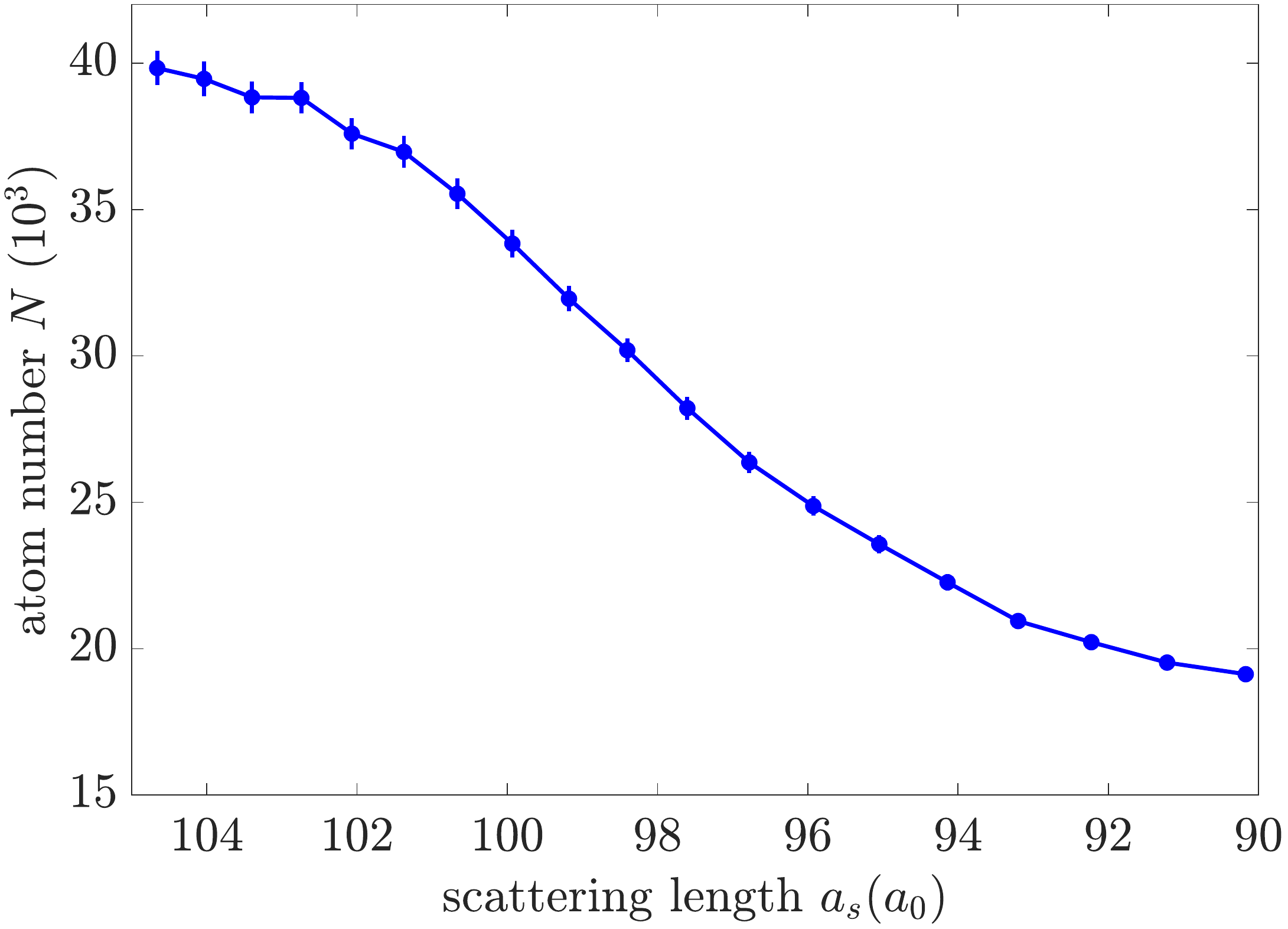}
	\caption{\textbf{Average atom number.} Average atom number for the scattering length range in the experiment. Crossing the phase transition from BEC to droplet arrays the increasing density leads to larger three-body losses causing lower atom numbers for smaller scattering lengths. Error bars indicate one standard deviation from the mean.}
	\label{fig:atomnumber}
\end{suppfigure}

We note that Ref.~\cite{Bottcher2019} has shown that the dynamical preparation scheme with the scattering length ramp only leads to states close to the actual ground state of the system for the BEC and supersolid regime. In regime of isolated droplets, the preparation process can lead to a different number of droplets from realization to realization. In comparison to the BdG theory, where only excitation on top of the actual ground state are considered, this could lead to increased fluctuations for the isolated droplets in the experiment. For all scattering lengths, we expect the dynamical nature of the sample preparation to slightly modify the observed fluctuations compared to purely thermally populated collective modes.

In order to get a better intuition of the different contributions to the experimental structure factor, we use a model-free statistical analysis, the principle component analysis (PCA) \cite{Jolliffe2002}. PCA has a wide range of general applications \cite{Jolliffe2002}, from image analysis to dimensional reduction of large dataset. For ultracold atomic systems, it turns out that there is a direct correspondence \cite{Dubessy2014} between the principal components (PCs) and the density variation of a mode obtained by the Bogoliubov-de-Gennes (BdG) formalism. This allows us to identify a specific PC with a certain BdG mode as long as the imaging noise is negligible.
One of PCA's properties is that the signal (in our case the centered images $n_j(\boldsymbol{r})$) can be reconstructed exactly using a superposition of all PCs, using their respective weights. However, a small subset of PCs accounts for most of the information contained within the scattering length scan. This becomes essential when PCA is used for dimensional reduction. 

In the experiment we combine all images independent of their scattering length to one large dataset. This allows us not only to illustrate the different contributions to the fluctuations and therefore the structure factor but also to track the weight of different PCs over a certain scattering length range \cite{Dubessy2014}. We confirm by treating the BEC and droplet regime separately that the relevant low-lying PCs do not change significantly except for their order (or variance). The first two PCs, namely the roton modes, do not change their shape over the complete scattering length range. By limiting the scattering length range to the BEC or droplet side we find in addition to the roton modes only the quadrupole or breathing modes, respectively. This is in agreement with the results presented in the main text, where we analyzed the complete dataset and observed a vanishing weight of these components on the respective side of the phase transition.

\subsection{Simulation details}
In this section, we briefly summarize simulation details not explained in the main text. A system of dipolar atoms that undergoes the transition from a BEC to a supersolid can be described by means of the extended Gross-Pitaevskii equation (eGPE) \cite{Ronen2006,Wenzel2017,Roccuzzo2019}

\begin{equation}\label{eq:GPE}
i \hbar \partial_t \psi = H_\mathrm{GP} \psi,
\end{equation}

\noindent where we define ${H_\mathrm{GP} \coloneqq H_0 + g|\psi|^2 + \Phi_\mathrm{dip}[\psi] + g_\mathrm{qf} |\psi|^3}$ and $\psi$ is normalized to the atom number ${N=\int \mathrm{d}^3r\, |\psi(\boldsymbol{r})|^2}$. The term ${H_0 = -\hbar^2 \nabla^2 / 2m + V_\mathrm{ext}}$ contains the kinetic energy and trap confinement ${V_\mathrm{ext}(\boldsymbol{r}) = m(\omega_x^2 x^2 + \omega_y^2 y^2 + \omega_z^2 z^2)/2}$.  The contact interaction strength $g = 4\pi\hbar^2a_\mathrm{s}/m$ is given by the scattering length $a_\mathrm{s}$. The dipolar mean field potential is $\Phi_\mathrm{dip} = \int \mathrm{d}^3r'\,  	V_\mathrm{dd}(\boldsymbol{r}-\boldsymbol{r}') |\psi(\boldsymbol{r}', t)|^2$ where $V_\mathrm{dd}(\boldsymbol{r}) = \frac{3g_\mathrm{dd}}{4\pi} \frac{1-3\cos^2 \vartheta}{|\boldsymbol{r}|^3}$ is the dipolar interaction for aligned dipoles. The strength of the dipolar interaction is given by the parameter $g_\mathrm{dd} = 4\pi\hbar^2a_\mathrm{dd}/m$ characterized by the dipolar length $a_\mathrm{dd} = \mu_0 \mu_\mathrm{m}^2 m / (12 \pi \hbar^2)$. Here, $\mu_\mathrm{m}$ is the magnetic moment and $\vartheta$ is the angle between $\boldsymbol{r}$ and the magnetic field axis. Furthermore, the quantity ${g_\mathrm{qf} = 32/(3\sqrt{\pi}) g a^{3/2} Q_5 (\epsilon_\mathrm{dd})}$ represents quantum fluctuations within the local density approximation for dipolar systems \cite{Lima2011,Lima2012}, where $\epsilon_\mathrm{dd} = g_\mathrm{dd} / g = a_\mathrm{dd} / a_\mathrm{s}$ is the relative dipolar strength. In our simulations, we use the approximation  ${Q_5(\epsilon_\mathrm{dd}) = 1+ \frac{3}{2}\epsilon_\mathrm{dd}^2}$ \cite{Lima2012,Ferrier-Barbut2016,Wenzel2017,Bisset2016,Baillie2017}. The mean-field dipolar potential is effectively calculated using a Fourier transform, where we use a spherical cutoff for the dipolar potential. The cutoff radius is set to the size of the simulations space such that there is no spurious interaction between periodic images \cite{Goral2002,Ronen2006,Lu2010}.

In order to study the elementary excitations, we use the BdG formalism as described in Ref.~\cite{Hertkorn2019} and linearly expand the wavefunction ${\psi(\boldsymbol{r},t) = \psi_0(\boldsymbol{r}) + \lambda \lbrack u(\boldsymbol{r}) e ^{-i\omega t} + v^{\ast}(\boldsymbol{r})e^{i\omega t}\rbrack e^{-i\mu t/\hbar}}$ around the ground state $\psi_0$ with the chemical potential $\mu$. This ansatz together with equation~\eqref{eq:GPE} leads to a system of linear equations that can be expressed in matrix form. For the actual form of the BdG matrix representation we refer the interested reader to the literature \cite{Baillie2017,Chomaz2018,Hertkorn2019}. We numerically solve these equations to obtain the modes $u$ and $v$ corresponding to the lowest excitation energies $\hbar \omega$. Due to our finite-sized system, we obtain a discrete spectrum of elementary excitations, rather than a continuous one. In addition there is a systematic shift between the theoretical and experimental transition point of approximately $\Delta a_s \simeq 2.6\,a_0$ \cite{Bottcher2019droplet,Petter2019,Chomaz2018,Natale2019,Petter2020highEnergyBragg}.

\subsection{Temperature-enhanced static structure factor}

In order to obtain the dynamic structure factor from the Bogoliubov-spectrum we first define the theoretical density fluctuation corresponding to the mode $j$ as ${\delta n_j = f_j^*\psi_0}$ with ${f_j=u_j+v_j}$ and the ground state $\psi_0$. The static structure factor then reads 

\begin{equation}\label{eq:SFDynTemp}
\begin{split}
S(\boldsymbol{k},\omega) &= \sum_j \left| \delta n_j(\boldsymbol{k})\right|^2  (( \bar{n}(\omega_j,T)+1)\delta(\omega-\omega_j)  \\
&+ \bar{n}(\omega_j,T)\delta(\omega+\omega_j) ),
\end{split}
\end{equation}

\noindent where $\delta n_j(\boldsymbol{k})$ is the Fourier transform of the fluctuations corresponding to the $j$'th mode and ${\bar{n}(\omega,T) = \left(\exp\left(\hbar\omega/k_\mathrm{B}T\right)-1\right)^{-1}}$ is the Bose-Einstein distribution. The static structure factor  $S(\boldsymbol{k}) = N^{-1} \int  \mathrm{d}\omega\, S(\boldsymbol{k}, \omega)$ is then given after integrating along $\omega$ \cite{PitaevskiiBook2016} which leads to the result 
\begin{equation}\label{eq:SFStatTemp}
S(\boldsymbol{k}) = N^{-1} \sum_j \left|\delta n_j(\boldsymbol{k})\right|^2\coth\left(\frac{\hbar\omega_j}{2k_\mathrm{B}T}\right),
\end{equation}
similar to equation~\eqref{eq:StrucTemp} of the main text. Here we made use of the identity ${2\bar{n}(\omega_j,T)+1=\coth(\hbar\omega_j/2k_\mathrm{B}T)}$. Equation~\eqref{eq:SFStatTemp} clearly indicates that low-lying modes satisfying the condition $\hbar \omega_j \ll k_\mathrm{B} T$ can be drastically enhanced. For our situation this means that the roton modes that soften towards the phase transition dominate the structure factor compared to all other modes. Only a finite number of modes are essentially necessary to quantitatively describe experiments at finite temperature.

\begin{suppfigure}[tb!]
	\includegraphics[trim=0 0 0 0,clip,scale=0.30]{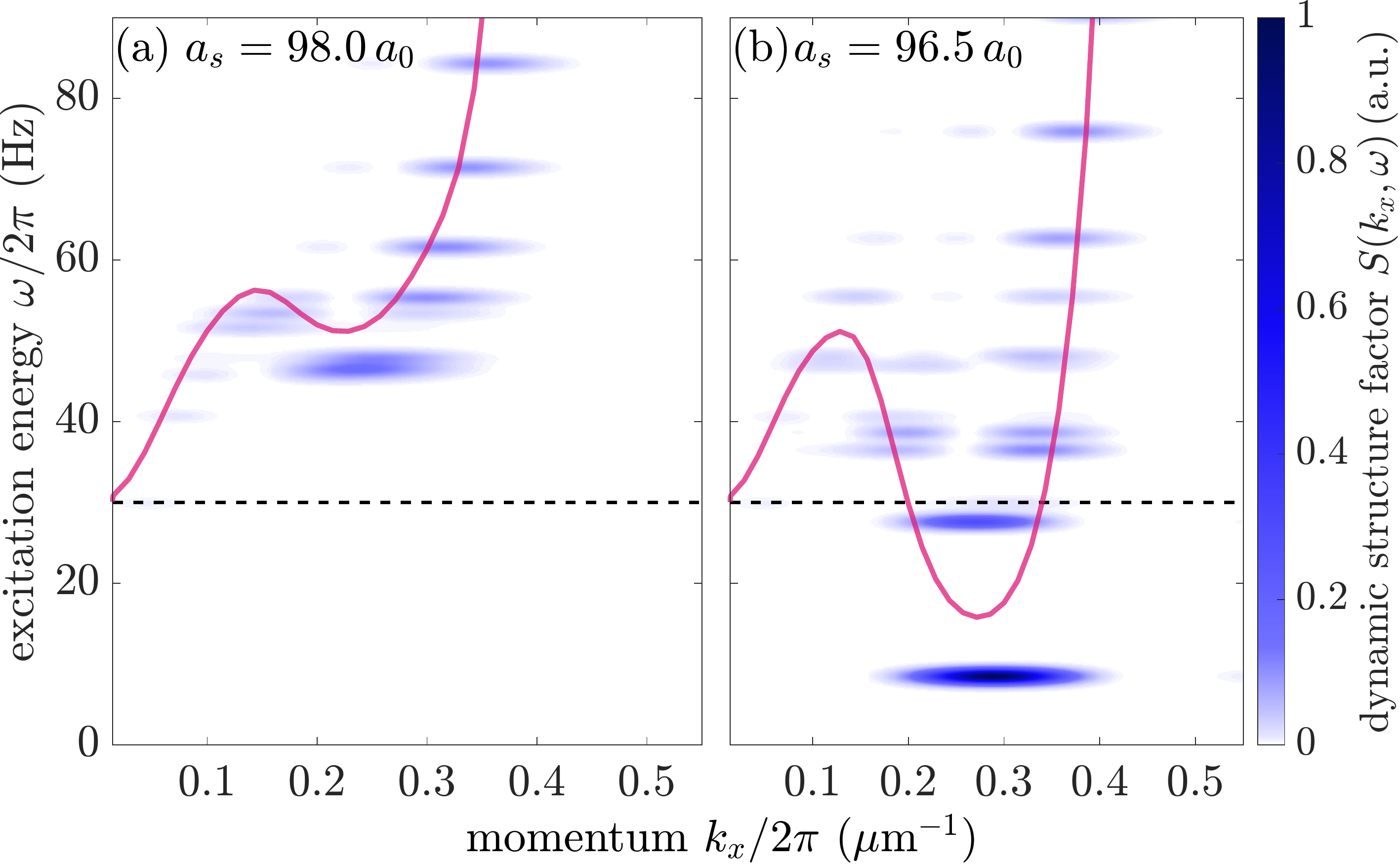}
	\caption{\textbf{Dynamic structure factor from numerical simulations.} Dynamic structure factor for two different scattering lengths before (a) and just after the transition (b). The finite system size results in discrete modes. The red line is calculated via the Feynman-Bijl formula equation~\eqref{eq:StrucTemp} at $T=0$. The softening roton mode clearly dominates the dynamic structure factor when approaching the phase transition. The colorbar shows the amplitude of the dynamic structure factor. For illustration purposes the dynamic structure factor was convolved along the $\omega$-axis with a Gaussian of width $\sigma = \SI{0.5}{\hertz}$.}
	\label{fig:SFDynamic}
\end{suppfigure}

In Fig.~\ref{fig:SFDynamic} we show the zero-temperature dynamic structure factor for two different scattering lengths, in the BEC phase and just after the phase transition. Due to our finite system size one can clearly see the discrete mode structure along the $k_x$- and $\omega$-axis. The color bar indicates the amplitude of the dynamic structure factor. The red line shows the excitation energy obtained via the Feynman-Bijl formula, equation~\eqref{eq:StrucTemp} of the main text at $T=0$. It illustrates that the continuous dispersion relation obtained from this equation yields a meaningful estimate of the discrete Bogoliubov spectrum.
\\
\subsection{Temperature dependence of the experimental excitation spectrum}
In Fig.~\ref{fig:Dispersion} of the main text we show the dispersion relation determined from the experimental static structure factor. This is done by solving equation~\eqref{eq:StrucTemp} numerically.  For dilute, weakly-interacting Bose gases with negligible quantum and thermal depletion this is a valid approximation. Although the system is undergoing a phase transition, the gas parameter $na_s^3\simeq 10^{-5}$ is still small enough to consider it as dilute and weakly-interacting with negligible quantum depletion, which is the case in our situation. The thermal component at a temperature of \SI{20}{\nano\kelvin} can be estimated to be less than 5\,\% \cite{PitaevskiiBook2016}.

In the situation where, as we have seen above, the structure factor is mainly dominated by the contribution from the two degenerate roton modes, it is possible to expand $\coth x  \simeq 1/x$ in equation~\eqref{eq:StrucTemp} for small energies \cite{Klawunn2011}, yielding
\begin{equation}\label{eq:StrucTempApprox}
S(\boldsymbol{k},T) \simeq S(\boldsymbol{k},0)\frac{2k_\mathrm{B}T}{\varepsilon(\boldsymbol{k})} = \frac{\hbar^2k^2k_\mathrm{B}T}{m\varepsilon(\boldsymbol{k})^2}.
\end{equation}
Equation~\eqref{eq:StrucTempApprox} might then also be used in a second step to calculate back the zero-temperature static structure factor. In fact we find due to the validity of this expansion that the expression $\epsilon(\boldsymbol{k}) \simeq \sqrt{\hbar^2 \boldsymbol{k}^2 k_\mathrm{B}T/mS(\boldsymbol{k},T)}$ is a good approximation to the numerical solution of equation~\eqref{eq:StrucTemp}.

The usage of equation~\eqref{eq:StrucTemp} requires knowledge of the temperature in the system. As measuring low temperatures with a vanishing thermal fraction is rather challenging, we show in Fig.~\ref{fig:SFTempError} how the uncertainty of our temperature of \SI{20 +- 5}{\nano\kelvin} affects the determined excitation spectrum for two scattering lengths.
\begin{suppfigure}[tb!]
	\includegraphics[trim=0 0 0 0,clip,scale=0.37]{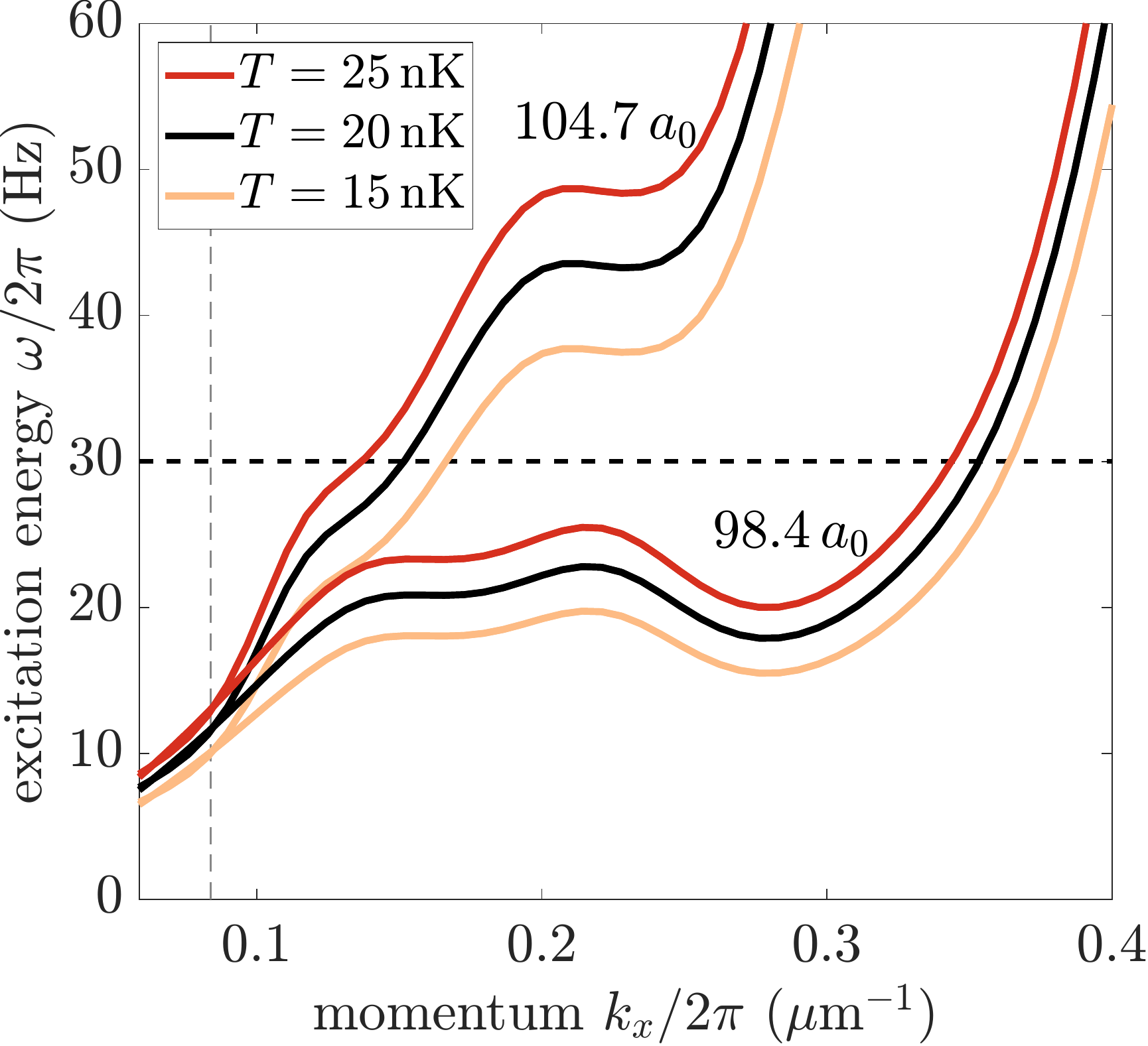}
	\caption{\textbf{Influence of assumed temperature on the excitation spectrum.} The assumed temperature influences the estimate obtained for the excitation spectrum based on equation~\eqref{eq:StrucTemp}. Examples of this influence are shown for a scattering length in the BEC regime ($104.7\,a_0$) and directly at the phase transition ($98.4\,a_0$).}
	\label{fig:SFTempError}
\end{suppfigure}

\end{document}